\documentclass{article}
\usepackage{latexsym,amsmath,amsopn,amssymb,amsthm,amsfonts}
\usepackage{graphicx} 
\usepackage{authblk}
\usepackage[T2A]{fontenc}
\usepackage[cp1251]{inputenc}
\usepackage{amsmath}
\usepackage[mathscr]{eucal}
\usepackage[english]{babel}
\usepackage[left=2.4cm, right=2cm, top=2cm, bottom=2.4cm]{geometry}
\usepackage{xcolor}

\title{Conditional Principle of Least Action: \\ Gauge Algebra and Noether's Second Theorem}
\author[]{V.~A.~Abakumova}
\author[]{S.~L.~Lyakhovich}
\affil[]{\small{Department of Quantum Field Theory, Physics Faculty, Tomsk State University, \\ Novosobornaya Sq. 1, Tomsk 634050, Russia}\protect\\
E-mail: \texttt{abakumova@phys.tsu.ru, sll@phys.tsu.ru}}

\date{} 

\begin{document}

\maketitle
\begin{abstract}
We consider the problem of a conditional extremum of an action within a class of fields constrained by differential equations. For this setup, we establish the interconnection between the gauge symmetry algebra of the equations constraining the admissible fields and that of the action restricted to this class of fields.  Based on this relation between the two gauge algebras, we derive an extension of Noether's second theorem and its converse. The general method is illustrated by the example of linearized gravity with the Einstein--Hilbert action and the class of admissible metrics restricted by the Cotton equation.
\end{abstract}
\maketitle

\section{Introduction}
Noether's second theorem and its converse establish a one-to-one correspondence between the gauge symmetry transformations of the action and the gauge identities between the corresponding Lagrange equations. Given the vast body of literature on Noether's theorems, we do not overview this area in this short article. For a general review and further literature, we refer to the book \cite{Kosmann:2010}. This connection between gauge symmetries and Noether identities is the starting point for constructing the gauge algebra of Lagrangian field theories. In turn, the gauge algebra is known to play an important role in field theory in many ways, including issues of quantization and the search for consistent interactions. For generalities of Lagrangian gauge systems, we refer to \cite{Henneaux:1992ig}.  For general non-Lagrangian field equations, there is no pairing between gauge symmetries and gauge identities. If the field equations admit certain auxiliary structures, a one-way correspondence is known to exist. The Lagrange anchor~\cite{Kazinski:2005eb} maps gauge identities to gauge symmetries, while the pre-symplectic structure~\cite{Crnkovic:1987,Havkine:2013} provides the inverse mapping~\cite{Grigoriev:2016wmk,Grigoriev:2021wgw,Sharapov:2016sgx,Sharapov:2016qne}. A similar picture can be seen with the connection between global symmetries and conserved currents established by the first Noether theorem: it is a one-to-one correspondence for Lagrangian systems, while the same auxiliary structures can provide a one-way mapping for the non-Lagrangian case. By introducing Lagrange multipliers, any set of equations can be made variational, so that Noether's theorems become valid. However, including Lagrange multipliers may increase the number of degrees of freedom, which will break the equivalence with the original system.   

In this work, we consider field theories whose equations of motion follow from the conditional principle of least action, when the extremum is sought in a class of field configurations constrained by a system of differential equations. In~\cite{Lyakhovich:2024yai}, conditionally Lagrangian equations were proposed that provide the necessary conditions for a conditional extremum of the action and do not include Lagrange multipliers or any other auxiliary fields. This system is non-Lagrangian as such, but it turns out to admit an extension of Noether's first theorem~\cite{Lyakhovich:2026eiu} that provides a one-to-one correspondence between global symmetries and conserved currents.   The subtle detail is that the conserved quantities turn out to be associated with action symmetries of a special type. Conversely, not every mutual symmetry of the constraints and action is connected with a conserved current -- only special symmetries are generated by conserved currents.  As we will demonstrate in this article, Noether's second theorem and its converse can be extended to the conditionally Lagrangian equations. Here, as in the case of Noether's first theorem, there are both obvious and less obvious differences from the unconditional extremum problem. The most obvious difference is that in the conditional extremum problem there is no pairing between fields and field equations, unlike in the unconditional extremum problem. Therefore, the gauge symmetry generators acting on the fields cannot immediately generate Noether identities between the field equations. We introduce an operator which we call \emph{the gauge symmetry reducer}. It reduces the gauge symmetry of the equations imposed to constrain the fields to the mutual gauge symmetry of the constraints and action. The formal Hermitian conjugate of the gauge symmetry reducer generates the Noether identities between the conditionally Lagrangian equations.
The less obvious distinction of the conditional extremum problem is that the Lagrangian is defined with a broader ambiguity than just a total divergence. In addition to terms vanishing by virtue of the constraints, gauge invariants of the system that constrains the fields can be added to the Lagrangian. Such additions do not affect the conditional extremals. The gauge transformations which change the Lagrangian by such invariants result in a gauge symmetry of the field  equations that is not connected to gauge identities. 
If the gauge symmetry of the complete system of field equations changes the Lagrangian at most by a total divergence modulo constraints, it results in gauge identities between the equations. Conversely, every Noether identity in the  conditional extremum problem results in a gauge symmetry that changes the Lagrangian at most by a total divergence in the class of admissible field configurations. 

It is a long-known fact that gauge symmetries for Lagrangian systems result in trivial conserved currents; for recent discussions of this issue we refer to~\cite{Baulieu:2024rfp,Barnich:2025jna,BarroseSa:2025zch,BarroseSa:2025uxe}. For conditionally Lagrangian equations, this fact, as we shall see, remains valid after some modifications. There can exist currents that are conserved only by virtue of the equations that constrain the admissible fields, regardless of the extremum of any action. We term these currents conditionally trivial even though their charges can depend on Cauchy data, unlike the off-shell conserved Noether currents connected with the gauge symmetry of the unconditional extremum of the action. For a conditional extremum, the gauge symmetry results in conditionally trivial currents. Thus, with the above reservations, Noether's second theorem and its converse are extended to the problem of the conditional extremum of an action.  

The article is organized as follows. In the next section, we consider the gauge symmetry of the conditional extremum problem. We begin with the gauge symmetry of the equations constraining the admissible fields. We deduce the equations for the conditional extremals by imposing the requirement that the gauge variation of the action has to vanish. This consideration follows~\cite{Lyakhovich:2024yai}, and we include this material to make the article self-contained.  In this section, we also address the issue of ambiguities in the Lagrangian in the conditional extremum problem. In Section 3, we study the gauge symmetry of the action which is defined for the fields constrained by a system of differential equations. This gauge symmetry of the action turns out to be a specialization of the gauge symmetry of the constraining equations. The specialization operators become the generating elements of the gauge algebra of the conditionally Lagrangian system.  In these terms, we formulate and prove the two theorems establishing a one-to-one correspondence between the gauge symmetries and Noether identities of conditionally Lagrangian systems. At the end of this section, we demonstrate the triviality of the Noether currents connected to the gauge symmetry of the conditionally Lagrangian system. 
Section 4 contains a simple example of a conditionally Lagrangian system which illustrates the main ingredients of the general formalism. Section 5 is a conclusion where we discuss the open issues, mostly related to the BV-BRST formalism for conditionally Lagrangian systems and their quantization.

\section{Field equations for a conditional extremum of the action}
Let us consider a set of fields $\phi^i (x)$ and constrain them by imposing a system of differential~equations 
\begin{equation}\label{T}
\displaystyle T_a(x,\phi,\partial\phi,\partial^2\phi,\dots) =0\,.
\end{equation}
Consider infinitesimal gauge transformations generated by finite-order differential operators,
\begin{equation}\label{GT}
\displaystyle \delta_\epsilon\phi^i=\hat{R}{}^i{}_\alpha\epsilon^\alpha\,, \qquad \hat{R}{}^i{}_\alpha=\sum_{k=0}^{k_\text{max}}\!\stackrel{(k)}{R}{}\!^{i\mu_k\ldots\mu_1}_{\alpha}(x,\phi,\partial\phi,\partial^2\phi,\ldots)\,
\partial_{\mu_k}\ldots\partial_{\mu_1}\,,
\end{equation}
where $\displaystyle \stackrel{(k)}{R}{}\!^{i\mu_k\ldots\mu_1}_{\alpha}$ are smooth functions of the fields and their derivatives, and the gauge parameters $\epsilon^\alpha(x)$ are arbitrary smooth functions\footnote{Hereinafter, indices are numbered from right to left: $\partial_{\mu_3}\ldots\partial_{\mu_1}\phi=\partial_{\mu_3}\partial_{\mu_2}\partial_{\mu_1}\phi$, $\partial_{\mu_1}\ldots\partial_{\mu_1}\phi=\partial_{\mu_1}\phi$, $\partial_{\mu_0}\ldots\partial_{\mu_1}\phi=\phi$. Summation over repeated indices is assumed, unless explicitly stated otherwise.}. These transformations are assumed to be symmetries of~(\ref{T}), 
\begin{equation}\label{GT0}
\displaystyle \delta_\epsilon T_a{}_{\big|T=0}=0\,, \qquad \forall\,\epsilon^\alpha\,. 
\end{equation}
Given the definition of gauge symmetry (\ref{GT0}), under appropriate assumptions regarding the regularity of the functions $T_a$, the variation $\delta_\epsilon T_a$  should be bilinear in $T$, $\epsilon$ and their derivatives,
\begin{equation}\label{U}
\displaystyle \delta_\epsilon T_a= \hat{U}_a(T,\epsilon)\,, 
\end{equation}
\begin{equation}\label{Uhat}
\displaystyle 
\hat{U}_a(T,\epsilon) =\sum_{k=0}^{k_\text{max}}\sum_{l=0}^{l_\text{max}}\!\stackrel{(k,l)}U{}\!\!_{a\alpha}^{\phantom{a}b\mu_k\ldots\mu_1\nu_l\ldots\nu_1}(x,\phi,\partial\phi,\partial^2\phi,\ldots)\, \partial_{\nu_l}\ldots\partial_{\nu_1}T_b \,\partial_{\mu_k}\ldots\partial_{\mu_1}\epsilon^\alpha\,.
\end{equation}
Throughout this article, we use notation analogous to~(\ref{Uhat})  for expressions multilinear in functions and their derivatives.

The gauge generators are defined by relation~(\ref{GT0}) with a natural ambiguity, i.e., modulo terms that vanish at~$T_a=0$.  
Such terms are considered trivial generators. Gauge transformations are considered equivalent if they differ by trivial terms.

The generating set $\{\hat{R}_\alpha\}$ of gauge transformations~(\ref{GT}) is assumed to be complete in the sense that any gauge transformation respecting the system~(\ref{T}) is in the span of the operators  $\hat{R}_\alpha$, 
\begin{equation}\label{complete}
\displaystyle \delta_{\mathcal{E}}\phi^i=\hat{G}{}^i\mathcal{E}\,, \quad \delta_{\mathcal{E}}T_a{}_{\big|T=0}=0\,, \quad \forall\, \mathcal{E}(x) \quad \Rightarrow \quad \exists\,\, \hat{g}^\alpha\,:  \,\hat{G}{}^i = \hat{R}{}^i{}_\alpha \hat{g}^\alpha\,.
\end{equation}

Given the completeness of the generating set, the commutator of two gauge transformations with independent arbitrary parameters $\epsilon_1^\alpha$ and $\epsilon_2^\beta$ should reduce (modulo terms that vanish for $T_a=0$) to a transformation from the same set with a new parameter $[\epsilon_1,\epsilon_2]$, which is a skew-symmetric bilinear combination of the original parameters and their derivatives 
\begin{equation}\label{bracket-epsilon}
\displaystyle [\epsilon_1,\epsilon_2]^\gamma=\hat{C}{}^\gamma(\epsilon_1,\epsilon_2)\, , \quad \hat{C}{}^\gamma(\epsilon_1,\epsilon_2)=-\,\hat{C}{}^\gamma(\epsilon_2,\epsilon_1)\,, 
\end{equation}
\begin{equation}\label{GAdef}
\displaystyle [\delta_{\epsilon_1},\delta_{\epsilon_2}]=\delta_{\epsilon_1}\delta_{\epsilon_2}-\delta_{\epsilon_2}\delta_{\epsilon_1}=\delta_{[\epsilon_1,\epsilon_2]} + 
\hat{E}(\epsilon_1, \epsilon_2, T)\, .
\end{equation}
Expanding the left-hand side requires applying the variation to the field-dependent coefficients $\stackrel{(k)}{R}$ of the gauge generators $\hat{R}$ defined in~(\ref{GT}). Substituting the explicit form of the transformations~(\ref{GT}) into~(\ref{GAdef}) and applying the chain rule, we obtain the following expression:
\begin{equation}
\displaystyle [\delta_{\epsilon_1},\delta_{\epsilon_2}]\phi^i=\sum\limits_{k=0}^{k_\text{max}} \sum_{k'=0}^{k'_\text{max}}\Big[\frac{\partial\!\stackrel{(k)}{R}{}_{\beta}^{i\mu_k\ldots\mu_1}}{\partial(\partial_{\lambda_{k'}}\ldots\partial_{\lambda_1}\phi^j)}\partial_{\lambda_{k'}}\ldots\partial_{\lambda_1}\big(\hat{R}^j{}_\alpha\epsilon_1^\alpha\big)\Big]\partial_{\mu_k}\ldots\partial_{\mu_1}\epsilon_2^\beta - (\epsilon_1 \leftrightarrow \epsilon_2)\,,
\end{equation}
where $k_\text{max}$ is the highest order of the derivatives of the gauge parameters, and $k'_\text{max}$ is the highest order of the derivatives of the fields in the field-dependent coefficients. Collecting terms with the same derivatives of the gauge parameters, we arrive at  the general structure of the commutator:
\begin{equation}\label{GA}
\displaystyle [\delta_{\epsilon_1},\delta_{\epsilon_2}]\phi^i=\hat{R}^i{}_\gamma \hat{C}^{\gamma}(\epsilon_1, \epsilon_2)+\hat{E}^{i}(\epsilon_1,\epsilon_2,T)\,.
\end{equation}

Now, let us consider the most general action functional 
\begin{equation}\label{S}
\displaystyle S[\phi(x)]=\int d^d x\, L(x,\phi, \partial\phi,\partial^2\phi,\ldots)\,,
\end{equation}
and take its unrestricted variation, which reduces to the Euler-Lagrange derivative and a total divergence, 
\begin{equation}\label{deltaS}
\displaystyle \delta S=\int d^dx\,\Big(\frac{\delta L}{\delta \phi^i}\delta\phi^i+\partial_{\mu}A^{\mu}\Big)\,,
\end{equation}
\begin{equation}\label{E-L-der}
\displaystyle \frac{\delta L}{\delta \phi^i}=\sum\limits_{m=0}^{m_\text{max}}(-1)^m\partial_{\mu_m}\ldots\partial_{\mu_1}\frac{\partial L}{\partial(\partial_{\mu_m}\ldots\partial_{\mu_1}\phi^i)}\,,
\end{equation}
\begin{equation}\label{Amu}
\displaystyle A^{\mu}=\sum\limits_{m=1}^{m_\text{max}}
\sum\limits_{m'=0}^{m_\text{max}-m}\Big[(-1)^{m'}\partial_{\nu_{m'}}\ldots\partial_{\nu_1}\frac{\partial L}{\partial(\partial_{\nu_{m'}}\ldots\partial_{\nu_1}\partial_{\mu_{m-1}}\ldots\partial_{\mu_1}
\partial_\mu\phi^i)}\Big]\partial_{\mu_{m-1}}\ldots\partial_{\mu_1}\delta\phi^i\,,
\end{equation}
where $m_{\text{max}}$ is the highest order of derivatives in the Lagrangian.

If we seek an unconditional extremum, this variation must vanish for arbitrary $\delta\phi^i$, which results in the Lagrange equations.
For a conditional extremum, the same variation~(\ref{deltaS}) must vanish not for arbitrary variations of the fields but for those that do not take the fields out of the admissible class constrained by equations~(\ref{T}). The gauge variation~(\ref{GT}),~(\ref{GT0}) of the action meets this condition, so it must vanish at a conditional extremum.  By integrating by parts, the gauge variation of the action is brought to the form
\begin{equation}\label{delta-epsilon-S}
\displaystyle \delta_\epsilon S=\int d^dx\,\Big(\epsilon^\alpha\hat{R}^{\dagger}{}_\alpha{}^{i}\frac{\delta L}{\delta\phi^i} + \partial_\mu (A_{\epsilon}^\mu+B_{\epsilon}^\mu)\Big)\,,
\end{equation}
where $\hat{R}^\dagger{}_\alpha{}^{i}$ is the formal Hermitian conjugate operator for the gauge generator $\hat{R}^i{}_\alpha$~(\ref{GT}),
\begin{equation}\label{R-dagger}
\displaystyle \hat{R}^{\dagger}{}_\alpha{}^{i} F =\sum_{k=0}^{k_\text{max}}(-1)^k\partial_{\mu_k}\ldots\partial_{\mu_1}\big(\!\!\stackrel{(k)}{R}{}\!^{i\mu_k\ldots\mu_1}_\alpha F\big)\,,
\end{equation}
for any smooth function $F$, $A_{\epsilon}^\mu$ has the form  (\ref{Amu}) with $\delta\phi^i$ defined by~(\ref{GT}),
\begin{equation}\label{Amu-epsilon}
\displaystyle A_{\epsilon}^{\mu}=\sum\limits_{m=1}^{m_\text{max}}
\sum\limits_{m'=0}^{m_\text{max}-m}\Big[(-1)^{m'}\partial_{\nu_{m'}}\ldots\partial_{\nu_1}\frac{\partial L}{\partial(\partial_{\nu_{m'}}\ldots\partial_{\nu_1}\partial_{\mu_{m-1}}\ldots\partial_{\mu_1}
\partial_\mu\phi^i)}\Big]\partial_{\mu_{m-1}}\ldots\partial_{\mu_1}\big(\hat{R}{}^i{}_\alpha\epsilon^\alpha\big)\,,
\end{equation}
and 
\begin{equation}\label{Bmu-epsilon}
\displaystyle B_{\epsilon}^{\mu}=\sum\limits_{k=0}^{k_\text{max}-1}\sum\limits_{k'=0}^{k_\text{max}-k-1}\Big[(-1)^{k'}\partial_{\nu_{k'}}\ldots\partial_{\nu_1}\Big(\frac{\delta L}{\delta \phi^i}\!\stackrel{(k'+1+k)}{R}{}\!_\alpha^{i\nu_{k'}\ldots\nu_1\mu\mu_k\ldots\mu_1}\Big)\Big]\partial_{\mu_k}\ldots\partial_{\mu_1}\epsilon^\alpha\,.
\end{equation}
Since the gauge variation~(\ref{delta-epsilon-S}) must vanish at the conditional extrema of the action, while the gauge parameters are arbitrary smooth functions, the conditional extremals necessarily obey the equations
\begin{equation}\label{Field-Eq}
\displaystyle \hat{R}^{\dagger}{}_\alpha{}^{i}\frac{\delta L}{\delta\phi^i}=0\,, \qquad T_a(x,\phi, \partial\phi, \partial^2\phi,\ldots)=0\,.
\end{equation}
These equations were first derived in~\cite{Lyakhovich:2024yai}, where it was also shown that any solution for the original fields of the Lagrange equations for the action with Lagrange multipliers obeys equations~(\ref{Field-Eq}). The system with Lagrange multipliers is generally weaker than~(\ref{Field-Eq}), since Lagrange multipliers can require independent initial data~\cite{Lyakhovich:2024yai}. While equations~(\ref{Field-Eq}) are not Lagrangian as such, they can be brought to a Hamiltonian form with a non-canonical Poisson bracket~\cite{Lyakhovich:2024yai}. The bracket is defined by the gauge generators and structure functions of the gauge algebra considered in this section.  As these field equations admit a Hamiltonian formalism, they can, in principle,  be quantized.  

To conclude this section, let us discuss the natural ambiguities of the Lagrangian in the conditional extremum problem.
The conditionally Lagrangian equations~(\ref{Field-Eq}) immediately follow from the requirement that the gauge variation~(\ref{GT})--(\ref{GT0}) of the action must vanish for the field configurations constrained by equations~(\ref{T}). Therefore, any two Lagrangians lead to the same conditional extrema if their gauge variations differ by a total divergence and terms that vanish for the fields obeying equations~(\ref{T}),
\begin{equation}\label{equivL1}
\displaystyle L'\,\sim\, L \quad \Leftrightarrow \quad \delta_\epsilon\left(L'\,-\, L \right)_{\big|T=0}=\text{total divergence}\,.
\end{equation}
In terms of the Lagrangians themselves, this ambiguity reads
\begin{equation}\label{equivL2}
\displaystyle L'\sim L \quad \Leftrightarrow \quad L'-L=I+\text{total divergence}\,, \quad \delta_\epsilon I_{\big| T=0}=\text{total divergence}\,.
\end{equation}
Any invariant of the gauge symmetry of equations~(\ref{T}) constraining the fields can be added to the Lagrangian without affecting the conditional extremals.
This invariant may include contributions that are combinations of $T$'s, but it does not necessarily reduce to the terms that vanish when $T=0$. Thus, the actions for equivalent Lagrangians~(\ref{equivL2}) do not coincide even for the fields obeying the constraint equations, but they result in the same conditionally Lagrangian equations. As we see, Lagrangians are more flexible in the conditional extremum problem than  in the unconditional one.  

\section{Gauge symmetry of a conditionally Lagrangian system \\ and Noether's second theorem}

Let us begin this section by discussing the gauge symmetry of the conditional extremum problem. We denote the gauge parameters of the entire conditionally Lagrangian system by~$\omega^A$, and the corresponding gauge generators by~$\hat{\Gamma}{}^i{}_A$,
\begin{equation}\label{Gamma}
\displaystyle \delta_\omega\phi^i=\hat{\Gamma}{}^i{}_A \omega^A\,, \qquad \hat{\Gamma}{}^i{}_A=
\sum_{q=0}^{q_\text{max}}\!\stackrel{(q)}{\Gamma}{}\!_A^{i\lambda_q\ldots\lambda_1} (x,\phi,\partial\phi,\partial^2\phi,\ldots)
\partial_{\lambda_q}\ldots\partial_{\lambda_1}\,.
\end{equation}
Clearly, the gauge symmetry transformations of the entire system must respect the equations constraining the trajectories, i.e., they should form a subset of the gauge transformations~(\ref{GT}),~(\ref{GT0}). Given the completeness of the generating set~(\ref{complete}), the generators $\hat{\Gamma}^i{}_A$~(\ref{Gamma}) must belong to the  span of the operators $\hat{R}^i{}_\alpha$~(\ref{GT}) that generate the gauge symmetry transformations of the equations $T_a=0$,
\begin{equation}\label{Gamma-rho}
\displaystyle \delta_\omega T_a{}_{\big| T=0}=0 \quad \Rightarrow \quad  \exists\,\,\hat{\rho}^\alpha{}_A\, :\, \hat{\Gamma}{}^i{}_A= \hat{R}{}^i{}_\alpha \hat{\rho}^\alpha{}_A \,,
\end{equation}
where $\hat{\rho}{}^\alpha{}_A$ is a differential operator with field-dependent coefficients, 
\begin{equation}\label{rho}
\displaystyle \hat{\rho}{}^\alpha{}_A=
\sum_{p=0}^{p_\text{max}}\!\stackrel{(p)}{\rho}{}\!_A^{\alpha\nu_p\ldots\nu_1} (x,\phi,\partial\phi,\partial^2\phi,\ldots)
\partial_{\nu_p}\ldots\partial_{\nu_1}\,,
\end{equation}
i.e., in~(\ref{Gamma})~$q_{\text{max}}=k_{\text{max}}+p_{\text{max}}$ (see~(\ref{GT}),~(\ref{rho})). The role of the operator $\hat{\rho}$ is to reduce the gauge symmetry of the equations constraining the admissible fields~(\ref{T}) to a gauge symmetry of the entire conditionally Lagrangian system.
For this reason,  we call the operator $\hat{\rho}{}^\alpha{}_A$ defined by relation~(\ref{Gamma-rho}) the \emph{\textbf{gauge symmetry reducer}} or the \emph{\textbf{gauge reducer}}, for brevity.
 
Given relation~(\ref{Gamma-rho}) between the gauge generators $\hat{R}^i{}_
\alpha$ of the constraint equations  and the gauge generators $\hat{\Gamma}^i{}_A$~(\ref{Gamma}) of the entire system,  the formal Hermitian conjugate operator $\hat{\Gamma}^\dagger{}_A{}^i$  reads
\begin{equation}\label{G-dag-R-dag}
\displaystyle \hat{\Gamma}^\dagger{}_A{}^i=\hat{\rho}{}^\dagger{}_A{}^\alpha\hat{R}{}^\dagger{}_\alpha{}^i\,.
\end{equation}

If we assume that the gauge symmetry transformations~(\ref{Gamma}) leave the entire system of equations~(\ref{Field-Eq}) invariant, then the Lagrangian may transform within the ambiguity allowed by the equivalence relations~(\ref{equivL2}),
\begin{equation}\label{GTLI}
\displaystyle \delta_\omega L = \omega^A I_A (x, \phi,\partial\phi, \partial^2\phi, \dots)+\text{total divergence}\,, \quad \delta_\epsilon I_A{}_{\big| T=0}=\text{total divergence}\,,
\end{equation}
where $\delta_\epsilon$ is the gauge symmetry transformation of equations~(\ref{T}). The transformed Lagrangian $L_\omega=L+\delta_\omega L$ results in a system equivalent to the original equations~(\ref{Field-Eq}). Thus, this transformation is a symmetry of the equations, while the Lagrangian changes not just up to a total divergence and terms vanishing at $T_a=0$.
If the invariants $I_A$ vanish at $T_a=0$,  the gauge transformation of the Lagrangian reads
\begin{equation}\label{GTL-proper}
\delta_\omega L = \omega^A \hat{W}{}^\dagger_A{}^a T_a+\text{total divergence}\,.
\end{equation}
We call transformations of this type a \emph{\textbf{proper gauge symmetry}} of a conditionally Lagrangian system. This symmetry respects not just the equations of motion, but also the Lagrangian, at least for the admissible configurations constrained by equations~(\ref{T}).
Given this definition of a proper gauge symmetry, the gauge generators $\hat{\Gamma}{}^i{}_A$ are defined with a natural ambiguity modulo on-shell vanishing terms. The corresponding ambiguity extends to the gauge reducer\footnote{A detailed description of these and similar ambiguities in the structures of the gauge algebra of conditionally Lagrangian systems will be discussed elsewhere, as it is not crucial for the main subject of this article.}.

\vspace{0.2 cm}
\noindent
\textbf{Theorem 1.} Every proper gauge symmetry of a conditionally Lagrangian system results in a Noether identity between the field equations.

\vspace{0.1 cm}

\noindent
\textbf{Proof.} 
Any variation of the action, including the infinitesimal gauge transformation (\ref{Gamma}), reduces to a combination of the Euler-Lagrange derivative and a total divergence, 
\begin{equation}\label{delta-omega-S}
\displaystyle \delta_\omega S=\int d^dx \Big(\frac{\delta L}{\delta \phi^i}\hat{\Gamma}{}^i{}_A\omega^A+\partial_\mu A_{\omega}^\mu\Big)\,,
\end{equation}
where 
\begin{equation}\label{Amu-omega}
\displaystyle A_{\omega}^{\mu}=\sum\limits_{m=1}^{m_\text{max}}
\sum\limits_{m'=0}^{m_\text{max}-m}\Big[(-1)^{m'}\partial_{\nu_{m'}}\ldots\partial_{\nu_1}\frac{\partial L}{\partial(\partial_{\nu_{m'}}\ldots\partial_{\nu_1}\partial_{\mu_{m-1}}\ldots\partial_{\mu_1}
\partial_\mu\phi^i)}\Big]\partial_{\mu_{m-1}}\ldots\partial_{\mu_1}\big(\hat{\Gamma}{}^i{}_A\omega^A\big)\,.
\end{equation}
By repeatedly integrating by parts in the first term in the variation~(\ref{delta-omega-S}), we reduce it to a form that includes two types of terms. The first group of terms includes the gauge parameters without derivatives, and the second group forms a total divergence,
\begin{equation}\label{dagger-Gamma}
\displaystyle \delta_\omega S= \int d^dx\,\Big(\omega^A\hat{\Gamma}{}^\dagger_A{}^i\frac{\delta L}{\delta \phi^i}+ \partial_\mu(A_{\omega}^\mu+B_{\omega}^\mu)\Big)\,,
\end{equation}
where $\hat{\Gamma}{}^\dagger_A{}^i$ is the formal Hermitian conjugate (\ref{G-dag-R-dag}) of the gauge generator~(\ref{Gamma}), and
$B_{\omega}^\mu$ reads 
\begin{equation}\label{Bmu-omega}
\displaystyle B_{\omega}^{\mu}=\sum\limits_{q=0}^{q_\text{max}-1}\sum\limits_{q'=0}^{q_\text{max}-q-1}\Big[(-1)^{q'}\partial_{\nu_{q'}}\ldots\partial_{\nu_1}\Big(\frac{\delta L}{\delta \phi^i}\!\stackrel{(q'+1+q)}{\Gamma}{}\!_\alpha^{i\nu_{q'}\ldots\nu_1\mu\mu_q\ldots\mu_1}\Big)\Big]\partial_{\mu_q}\ldots\partial_{\mu_1}\omega^A\,.
\end{equation}
The equality~(\ref{dagger-Gamma}) is valid for any action and any infinitesimal gauge transformation, be it a symmetry or not. 

A proper gauge symmetry of the system means that the Lagrangian is invariant under the transformations~(\ref{Gamma}) modulo a total divergence and terms vanishing at $T_a=0$ (see~(\ref{GTL-proper})). This means that the action variation~(\ref{delta-omega-S}) can be reduced to a total divergence and a term that is bilinear in the constraints $T_a$ (\ref{T}), the gauge parameters~$\omega^A$, and their derivatives, i.e., to  
\begin{equation}\label{LambdaW}
\displaystyle \delta_\omega S=\int d^dx\, \Big(\partial_\mu\Lambda^\mu+\hat{W}(T,\omega)\Big)\,, \qquad \hat{W}(T,\omega) =\sum_{r=0}^{r_\text{max}}\sum_{l=0}^{l_\text{max}}\!\stackrel{(r,l)}W{}\!_{A}^{a\mu_r\ldots\mu_1\nu_l\ldots\nu_1}\, \partial_{\nu_l}\ldots\partial_{\nu_1}T_a \,\partial_{\mu_r}\ldots\partial_{\mu_1}\omega^A\,,
\end{equation}
where $r_{\text{max}}=m_{\max}+q_{\text{max}}$. By repeatedly integrating (\ref{LambdaW}) by parts, we reduce it to terms with gauge parameters without derivatives and a total divergence~(cf.~(\ref{GTL-proper})),
\begin{equation}\label{delta-omega-S-2}
\displaystyle \delta_\omega S=\int d^dx\,\Big(\omega^A\hat{W}^\dagger_A{}^aT_a+\partial_\mu(\Lambda^\mu+K_{\omega}^\mu)\Big)\,,
\end{equation}
where
\begin{equation}
\displaystyle \hat{W}^\dagger_A{}^aT_a=\sum\limits_{r=0}^{r_\text{max}}\sum\limits_{l=0}^{l_\text{max}}(-1)^r\partial_{\mu_r}\ldots\partial_{\mu_1}\Big[\!\stackrel{(r,l)}W{}\!_{A}^{a\mu_r\ldots\mu_1\nu_l\ldots\nu_1}\, \partial_{\nu_l}\ldots\partial_{\nu_1}T_a\Big]\,,
\end{equation}
i.e.,
\begin{equation}
\hat{W}^\dagger_A{}^a=\sum\limits_{s=0}^{r_\text{max}+l_{\text{max}}}\!\!\sum\limits_{t=\text{max}(0,s-l_{\text{max}})}^{\text{min}(s,r_{\text{max}})}\sum\limits_{t'=t}^{r_\text{max}}\frac{(-1)^{t'}t'!}{t!(t'-t)!}\partial_{\mu_{t'}}\ldots\partial_{\mu_{t+1}}\Big(\!\stackrel{(t',s-t)}W{}\!_{A}^{a\mu_{t'}\ldots\mu_1\nu_{s-t}\ldots\nu_1}\Big)\,\partial_{\mu_t}\ldots\partial_{\mu_1}\partial_{\nu_{s-t}}\ldots\partial_{\nu_1}\,,
\end{equation}
and
\begin{equation}
\displaystyle K_\omega^\mu=\sum\limits_{r=0}^{r_\text{max}-1}\sum\limits_{r'=0}^{r_\text{max}-r-1}\sum\limits_{l=0}^{l_\text{max}}(-1)^{r'}\partial_{\lambda_{r'}}\ldots\partial_{\lambda_{1}}\Big[\!\stackrel{(r'+1+r,l)}W{}\!_{A}^{a\lambda{r'}\ldots\lambda_1\mu\mu_{r}\ldots\mu_1\nu_l\ldots\nu_1}\, \partial_{\nu_l}\ldots\partial_{\nu_1}T_a\Big]\partial_{\mu_{r}}\ldots\partial_{\mu_1}\omega^A\,.
\end{equation}

The two variations,~(\ref{dagger-Gamma})~and~(\ref{delta-omega-S-2}), must coincide since they are evaluated for the same action functional and with respect to the same gauge transformation of the fields. This leads to the identity
\begin{equation}\label{Int-identity}
\displaystyle \int d^dx\,\omega^A\Big(\hat{\Gamma}{}^\dagger_A{}^i\frac{\delta L}{\delta \phi^i} -\hat{W}^\dagger_A{}^aT_a\Bigg) = \int d^dx \partial_\mu\big(\Lambda^\mu+K_{\omega}^\mu-A_{\omega}^\mu-B_\omega^{\mu})\,.
\end{equation}
Since the gauge parameters $\omega^A$ are arbitrary smooth functions, they can be chosen to have compact support. As a result, the right-hand side of~(\ref{Int-identity}) vanishes by Stokes'~theorem. Then, by virtue of the fundamental lemma of the calculus of variations, we arrive at the identities between the conditionally Lagrangian equations,
\begin{equation}\label{NI-strong}
\displaystyle \hat{\rho}{}^\dagger{}_A{}^\alpha\hat{R}{}^\dagger{}_\alpha{}^i\frac{\delta L}{\delta\phi^i}-\hat{W}^\dagger_A{}^aT_a \equiv 0 .
\end{equation}
These relations are the Noether identities between the conditionally Lagrangian field equations~(\ref{Field-Eq}).
\vspace{0.1 cm}

Q.E.D.
\vspace{0.1 cm}

As we see, the proper gauge symmetry of the action results in relations~(\ref{NI-strong}), where the conjugate of the gauge reducer $\hat{\rho}^\dagger{}_A{}^\alpha$ acts as the Noether identity generator. 

If the gauge symmetry is weakened to the transformation~(\ref{equivL2}), then the field equations remain intact. This relaxed symmetry leads to weaker differential consequences of the field equations than the Noether identities,
\begin{equation}\label{NI-I}
\displaystyle \hat{\rho}{}^\dagger{}_A{}^\alpha\hat{R}{}^\dagger{}_\alpha{}^i\frac{\delta L}{\delta\phi^i}-\hat{W}^\dagger_A{}^aT_a \equiv I_A ,
\end{equation}
where $I_A$ is the gauge invariant involved in the transformation of the Lagrangian~(\ref{equivL2}). This relation means that the local quantity  $I_A$, being non-trivial at $T_a=0$, vanishes-on shell with respect to the complete system of field equations (\ref{Field-Eq}). At the moment, we do not have any reasonable example of such a relaxed symmetry, so the interpretation of these hypothetical consequences remains unclear.

\vspace{0.2 cm}

\textbf{Theorem 2.} If the field equations of a conditionally Lagrangian system satisfy a set of Noether identities, the underlying action and the equations constraining the trajectories are invariant under proper gauge transformations. 

The theorem is proved by reversing the line of reasoning in the proof of Theorem 1, starting from the Noether identities~(\ref{NI-strong}) and arriving at the proper gauge symmetry of the action~(\ref{Gamma}),~(\ref{Gamma-rho}),~(\ref{GTL-proper}).

Let us make a remark on the Noether identities (\ref{NI-strong}). The system of equations (\ref{T}) may admit its own differential identities that hold irrespective of any action,
\begin{equation}\label{II}
  \hat{N}{}^a T_a\equiv 0 \, .
\end{equation}
These internal identities of the constraining equations correspond, as is seen from~(\ref{NI-strong}), to a trivial gauge reducer~$\hat{\rho}{}^\alpha{}_A$.  The internal identities of system~(\ref{T}) are unrelated to the gauge symmetry of the action. If the gauge symmetry reducer $\hat{\rho}{}^\alpha{}_A$ is non-trivial on-shell, its formal Hermitian conjugate generates a Noether identity. 

Let us mention the currents that are conserved only by virtue of equations~(\ref{T}) constraining the admissible fields, irrespective of any action:
\begin{equation}\label{T-currents}
\displaystyle \partial_\mu\Theta^\mu=\hat{\theta}{}^a T_a\,.
\end{equation}
We call these currents \emph{\textbf{conditionally trivial}}.

Let us now consider the Noether currents corresponding to the identities (\ref{NI-strong}). The integrands of (\ref{delta-omega-S}), (\ref{LambdaW}) satisfy the following relation:
\begin{equation}
\displaystyle \sum_{p=0}^{p_\text{max}}\!\Big(\hat{R}^\dagger{}_\alpha{}^i\frac{\delta L}{\delta \phi^i}\Big)\!\stackrel{(p)}{\rho}{}\!_A^{\alpha\mu_p\ldots\mu_1}
\partial_{\mu_p}\ldots\partial_{\mu_1}\omega^A-\sum\limits_{r=0}^{r_\text{max}}\!\stackrel{(r)}W{}\!_{A}^{\mu_r\ldots\mu_1}\partial_{\mu_r}\ldots\partial_{\mu_1}\omega^A=\partial_\mu J^\mu\,,
\end{equation}
where 
\begin{equation}\label{Jmu}
\displaystyle J^\mu=\Lambda^\mu-A_\omega^\mu-\!\sum\limits_{k=0}^{k_\text{max}-1}\sum\limits_{k'=0}^{k_\text{max}-k-1}\Big[(-1)^{k'}\partial_{\nu_{k'}}\ldots\partial_{\nu_1}\Big(\frac{\delta L}{\delta \phi^i}\!\stackrel{(k'+1+k)}{R}{}\!_\alpha^{i\nu_{k'}\ldots\nu_1\mu\mu_k\ldots\mu_1}\Big)\Big]\partial_{\mu_k}\ldots\partial_{\mu_1}\big(\hat{\rho}^\alpha{}_A\omega^A\big)\,,
\end{equation}
and we use the notation
\begin{equation}
\displaystyle \stackrel{(r)}W{}\!_{A}^{\mu_r\ldots\mu_1}=\sum_{l=0}^{l_\text{max}} \!\stackrel{(r,l)}W{}\!_{A}^{a\mu_r\ldots\mu_1\nu_l\ldots\nu_1}\, \partial_{\nu_l}\ldots\partial_{\nu_1}T_a\,.
\end{equation}
The current $J^\mu$ is of order $j_\text{max}=r_\text{max}-1$ in the derivatives of the parameters $\omega^A$ of the transformations~(\ref{Gamma}),
\begin{equation}\label{Jmu-J}
\displaystyle J^\mu=\sum\limits_{j=0}^{j_\text{max}}\!\stackrel{(j+1)}J{}
\!_{A}^{\mu\mu_j\ldots\mu_1}\partial_{\mu_j}\ldots\partial_{\mu_1}\omega^A\,.
\end{equation}
The functions $\stackrel{(j+1)}J{}\!_{A}^{\mu\mu_j\ldots\mu_1}$ are symmetric in all their indices, except for the first one, and one can split them into the totally symmetric part and the remaining part,
\begin{equation}\label{J-2}
\displaystyle \stackrel{(j+1)}J{}\!_{A}^{\mu\mu_j\ldots\mu_1}=\!\stackrel{(j+1)}J{}\!_{A}^{(\mu\mu_j\ldots\mu_1)}+\!\stackrel{(j+1)}\Omega{}\!_{A}^{\mu\mu_j\ldots\mu_1}\,, \qquad \stackrel{(j+1)}\Omega{}\!_{A}^{(\mu\mu_j\ldots\mu_1)}\!=0\,.
\end{equation}
Following a procedure analogous to that described in~\cite{BarroseSa:2025zch}, we can represent the current $J^\mu$ in the form
\begin{equation}\label{Jmu-3}
\displaystyle J^\mu=\Psi^\mu+\Omega^\mu+\Theta^\mu\,,
\end{equation}
where the quantity
\begin{equation}\label{Psimu}
\displaystyle \Psi^\mu=\sum\limits_{j=0}^{j_\text{max}}\sum\limits_{j'=0}^{p_\text{max}-1-j}(-1)^{j'}\partial_{\nu_{j'}}\ldots\partial_{\nu_1}\Big[\Big(\hat{R}^\dagger{}_\alpha{}^i\frac{\delta L}{\delta \phi^i}\Big)\!\stackrel{(j'+j+1)}{\rho}{}\!_A^{\alpha\nu_{j'}\ldots\nu_1\mu\mu_j\ldots\mu_1}\Big]\partial_{\mu_j}\ldots\partial_{\mu_1}\omega^A
\end{equation}
vanishes on-shell, the divergence of the quantity
\begin{equation}\label{Omegamu}
\displaystyle \Omega^\mu=-\,\sum\limits_{j=0}^{j_\text{max}}\sum\limits_{j'=0}^{j_\text{max}-j}(-1)^{j'}\partial_{\nu_{j'}}\ldots\partial_{\nu_1}\Big(\!\stackrel{(j'+j+1)}{\Omega}{}\!_A^{\nu_{j'}\ldots\nu_1\mu\mu_j\ldots\mu_1}\Big)\partial_{\mu_j}\ldots\partial_{\mu_1}\omega^A
\end{equation}
vanishes off-shell, since
\begin{equation}
\begin{array}{c}
\displaystyle \partial_\mu\Omega^\mu=-\,\sum\limits_{j=1}^{j_\text{max}}\sum\limits_{j'=0}^{j_\text{max}-j}(-1)^{j'}\partial_{\nu_{j'}}\ldots\partial_{\nu_1}\partial_\mu\Big(\!\stackrel{(j'+j+1)}{\Omega}{}\!_A^{\nu_{j'}\ldots\nu_1\mu\mu_j\ldots\mu_1}\Big)\partial_{\mu_j}\ldots\partial_{\mu_1}\omega^A\\[3mm]
\displaystyle -\,\sum\limits_{j=0}^{j_\text{max}-1}\sum\limits_{j'=1}^{j_\text{max}-j}(-1)^{j'}\partial_{\nu_{j'}}\ldots\partial_{\nu_1}\Big(\!\stackrel{(j'+j+1)}{\Omega}{}\!_A^{\nu_{j'}\ldots\nu_1\mu\mu_j\ldots\mu_1}\Big)\partial_\mu\partial_{\mu_j}\ldots\partial_{\mu_1}\omega^A\equiv 0\,,
\end{array}
\end{equation}
and the quantity 
\begin{equation}\label{Thetamu}
\displaystyle \Theta^\mu=-\,\sum\limits_{j=0}^{j_\text{max}}\sum\limits_{j'=0}^{j_\text{max}-j}(-1)^{j'}\partial_{\nu_{j'}}\ldots\partial_{\nu_1}\Big(\!\stackrel{(j'+j+1)}{W}{}\!_A^{\nu_{j'}\ldots\nu_1\mu\mu_j\ldots\mu_1}\Big)\partial_{\mu_j}\ldots\partial_{\mu_1}\omega^A
\end{equation}
is conditionally trivial as it is a combination of constraints $T_a$ and their derivatives (see~(\ref{T-currents})). In this way, decomposition~(\ref{Jmu-3})--(\ref{Thetamu}) generalizes the standard structure of improper gauge currents to conditionally Lagrangian systems: besides the part~$\Psi^\mu$ vanishing on the equations of conditional extremum and the identically conserved term~$\Omega^\mu$, the current~$J^\mu$ acquires an extra component~$\Theta^\mu$ that vanishes by virtue of the constraints $T_a=0$.

\section{Example: Linearized gravity constrained by the Cotton equation}

Let us illustrate the general formalism developed in Sections~2~and~3 by the example of linearized gravity in $d=4$ Minkowski space. The role of the fields $\phi^i$ is played by a symmetric second-rank tensor $h^{\mu\nu}(x)$. We consider the conditional extremum problem for the linearized Einstein--Hilbert action (cf.~(\ref{S})):
\begin{equation}\label{EH-action}
\displaystyle S=\frac{1}{4}\int d^4x \Big(\partial_\mu h_{\nu\lambda}\partial^\mu h^{\nu\lambda}+2\partial_\mu h\partial_\nu h^{\mu\nu}-2\partial^\mu h_{\nu\lambda}\partial^\lambda h^{\nu\mu}-\partial_\mu h\partial^\mu h\Big)\,, 
\end{equation}
where $\eta_{\mu\nu}=\text{diag}(1,-\,1,-\,1,-\,1)$ is the Minkowski metric and  $h\equiv \eta_{\mu\nu}h^{\mu\nu}$.

Assume that the class of admissible fields is constrained by the linearized Cotton equation, which plays the role of the system of differential equations $T_a=0$~(\ref{T}):
\begin{equation}\label{Cotton}
\begin{array}{c}
\displaystyle C_{\mu\nu\lambda}\equiv\frac{1}{2}\Big[\partial_\mu\partial^\rho(\partial_\lambda h_{\rho\nu}-\partial_\nu h_{\rho\lambda})-\square(\partial_\lambda h_{\mu\nu}-\partial_\nu h_{\mu\lambda})\Big]\\[3mm]
\displaystyle +\,\frac{1}{6}\Big[\eta_{\mu\lambda}\partial_\nu(\partial^\rho\partial^\sigma h_{\rho\sigma}-\square h)-\eta_{\mu\nu}\partial_\lambda(\partial^\rho\partial^\sigma h_{\rho\sigma}-\square h)\Big]=0\,.
\end{array}
\end{equation}
The Cotton tensor is invariant (with $\hat{U}_a=0$ in~(\ref{U})) under the infinitesimal linearized diffeomorphisms and Weyl transformations~(cf.~(\ref{GT0})):
\begin{equation}\label{GT-C}
\delta_\varepsilon C_{\mu\nu\lambda}\equiv 0\,, \qquad \delta_{\varepsilon} h^{\mu\nu}=\partial^\mu\varepsilon^\nu+\partial^\nu\varepsilon^\mu+\eta^{\mu\nu}\varepsilon\,,
\end{equation}
i.e., the gauge generator $\hat{R}^i{}_\alpha$~(\ref{GT}) can be split into two parts corresponding to the independent gauge parameters $\varepsilon^\alpha(x)$ describing local space-time shifts and the scalar parameter $\varepsilon(x)$ related to dilations: 
\begin{equation}\label{GT-Cotton}
\displaystyle \delta_{\varepsilon} h^{\mu\nu}=\hat{R}^{\mu\nu}{}_\alpha\varepsilon^\alpha+R^{\mu\nu}\varepsilon\,, \qquad \hat{R}^{\mu\nu}{}_\alpha=\delta^\nu_\alpha\partial^\mu+\delta^\mu_\alpha\partial^\nu\,, \qquad R^{\mu\nu}=\eta^{\mu\nu}\,.
\end{equation}

According to~(\ref{delta-epsilon-S})--(\ref{Field-Eq}), the conditionally Lagrangian equations are obtained by applying the formal Hermitian conjugates of the gauge generators~(\ref{GT-Cotton}), 
\begin{equation}
\displaystyle \hat{R}^{\dagger}{}_\alpha{}^{\mu\nu}=-\delta^\nu_\alpha\partial^\mu-\delta^\mu_\alpha\partial^\nu\,, \qquad \hat{R}^{\dagger}{}^{\mu\nu}=\eta^{\mu\nu}\,,
\end{equation}
to the Euler-Lagrange derivatives~(\ref{E-L-der}) of the Lagrangian,
\begin{equation}\label{EH-eq}
\displaystyle \frac{\delta L}{\delta h^{\mu\nu}}\equiv\frac{1}{2}\big(\partial_\mu\partial^\lambda h_{\nu\lambda}+\partial_\nu\partial^\lambda h_{\mu\lambda}-\square h_{\mu\nu}-\partial_\mu\partial_\nu h\big)-\frac{1}{2}\eta_{\mu\nu}\big(\partial^\lambda\partial^\rho h_{\lambda\rho}-\square h\big)\,.
\end{equation} 
This yields two sets of equations corresponding to the variations with respect to $\varepsilon^\alpha$ and $\varepsilon$:
\begin{equation}\label{Field-Eq-LG}
\displaystyle \frac{\delta S}{\delta \varepsilon^\alpha}=\hat{R}^{\dagger}{}_\alpha{}^{\mu\nu}\frac{\delta L}{\delta h^{\mu\nu}}\equiv-\,2\partial^\mu\frac{\delta L}{\delta h^{\mu\alpha}}=0\,,  \qquad 
\frac{\delta S}{\delta \varepsilon}=\hat{R}^{\dagger}{}{}^{\mu\nu}\frac{\delta L}{\delta h^{\mu\nu}}\equiv \square h-\partial^\mu\partial^\nu h_{\mu\nu}=0\,.
\end{equation}
Note that for the unconditional extremum one would have ${\delta L}/{\delta h^{\mu\nu}}=0$. The scalar equation in~(\ref{Field-Eq-LG}), which expresses the vanishing of the linearized scalar curvature, is nothing other than the linearized Nordstr\"om equation.

Now let us consider the proper gauge symmetry of the entire conditionally Lagrangian system~(\ref{Cotton}),~(\ref{Field-Eq-LG}). While the constraining equation~(\ref{Cotton}) is invariant under both diffeomorphisms and Weyl transformations~(\ref{GT-Cotton}), the action~(\ref{EH-action}) is invariant only under the linearized diffeomorphisms (so that $\hat{W}=0$ in~(\ref{LambdaW})):
\begin{equation}\label{GT-condLG}
\displaystyle \delta_{\omega} S=0\,, \qquad \delta_{\omega} C_{\mu\nu\lambda}=0\,: \qquad \delta_\omega h^{\mu\nu}=\partial^\mu\omega^\nu+\partial^\nu\omega^\mu\,.
\end{equation}
Comparing~(\ref{GT-condLG}) with the general form of the gauge transformations of the complete system~(\ref{Gamma}), we identify the gauge generators $\hat{\Gamma}^{\mu\nu}{}_\alpha$:
\begin{equation}
\displaystyle \delta_{\omega} h^{\mu\nu}=\hat{\Gamma}^{\mu\nu}{}_\alpha{\omega}^\alpha\,, \qquad \hat{\Gamma}^{\mu\nu}{}_\alpha=\delta^\nu_\alpha\partial^\mu+\delta^\mu_\alpha\partial^\nu\,.
\end{equation}
Using the general relation~(\ref{Gamma-rho}) between the gauge generators~$\hat{\Gamma}$ of the entire system and  the gauge generators~$\hat{R}$ of the constraints, we find the explicit form of the gauge symmetry reducer~(\ref{rho}), which consists of two components, $\hat{\rho}^\beta{}_\alpha$ and $\hat{\rho}_\alpha$:
\begin{equation}\label{rho-LG}
\displaystyle \hat{\Gamma}^{\mu\nu}{}_\alpha=\hat{R}^{\mu\nu}{}_\beta\hat{\rho}^\beta{}_\alpha+\hat{R}^{\mu\nu}\hat{\rho}_\alpha \qquad \Rightarrow \qquad \hat{\rho}^\beta{}_\alpha=\delta^\beta_\alpha\,, \qquad \hat{\rho}_\alpha=0\,.
\end{equation}

By virtue of Theorem~1 in Section~3, substituting the formal Hermitian conjugate of the gauge reducer~(\ref{rho-LG}) into the general expression~(\ref{NI-strong}) (with $\hat{W}^\dagger=0$), we arrive at the Noether identities for the conditionally Lagrangian equations~(\ref{Field-Eq-LG}):
\begin{equation}\label{NI-LG}
\displaystyle \hat{\rho}^\dagger{}_\alpha{}^{\beta}\hat{R}^{\dagger}{}_\beta{}^{\mu\nu}\frac{\delta L}{\delta h^{\mu\nu}}\equiv0\,,  \qquad \hat{\rho}^{\dagger}{}_\alpha{}^{\beta}=\delta^\beta_{\alpha}\,.
\end{equation}
As we see from~(\ref{rho-LG})~and~(\ref{NI-LG}), since the diffeomorphism symmetry of the constraint~(\ref{Cotton})  is preserved by the action~(\ref{EH-action}), the gauge reducer in this sector is the identity operator,  $\hat{\rho}^\beta{}_\alpha=\delta^\beta_\alpha$. Consequently, the Noether identities~(\ref{NI-LG}) coincide with the first set of the conditionally Lagrangian equations in~(\ref{Field-Eq-LG}), ${\delta S}/{\delta \varepsilon^\alpha=0}$, which turn out to be satisfied identically off-shell due to the contracted Bianchi identities for the linearized Einstein tensor~(\ref{EH-eq}). Conversely, the Weyl transformations in~(\ref{GT-C}) are not a symmetry of the action~(\ref{EH-action}), so the corresponding component of the gauge reducer vanishes, $\hat{\rho}_\alpha=0$. As a result, the Weyl variation in~(\ref{Field-Eq-LG}), $\delta{S}/\delta{\varepsilon}=0$, is not subject to any Noether identity and survives as the only non-trivial conditionally Lagrangian equation. Thus, the conditional extremum problem for the Einstein-Hilbert action~(\ref{EH-action}) consistently reduces to the joint system of the linearized Cotton and Nordstr\"om equations. This joint system was previously obtained in~\cite{Lyakhovich:2024yai} by extremizing the linearized Eistein-Hilbert action in the class of metrics constrained by the linearized Nordsrt\"om equation. As demonstrated in~\cite{Lyakhovich:2024yai}, this system describes 6 local degrees of freedom by the phase-space count (compared to 4 for the linearized Einstein equations), with the 2 additional degrees of freedom corresponding to a massless scalar field coupled to linearized gravity.

Let us point out that the constraining Cotton equation~(\ref{Cotton}) also satisfies its own internal differential identities of the form~(\ref{II}), which hold off-shell irrespective of any action:
\begin{equation}\label{Cotton-diff}
\displaystyle \partial^\mu C_{\mu\nu\lambda}\equiv 0\,, \qquad \partial_\sigma C_{\mu\nu\lambda}+\partial_\nu C_{\mu\lambda\sigma}+\partial_\lambda C_{\mu\sigma\nu}\equiv0\,,
\end{equation}
as well as the algebraic identities
\begin{equation}\label{Cotton-alg}
\displaystyle C_{\mu(\nu\lambda)}\equiv 0\,, \qquad C_{[\mu\nu\lambda]}\equiv0\,, \qquad \eta^{\mu\nu}C_{\mu\nu\lambda}\equiv 0\,.
\end{equation}
As we discussed in Section~3, these internal identities correspond to a trivial gauge reducer and are unrelated to the proper gauge symmetry~(\ref{GT-condLG}) of the action.

Finally, let us write down  the explicit expression for the Noether current~$J^\mu$~(\ref{Jmu}) associated with the proper gauge symmetry~(\ref{GT-condLG}) of the action~(\ref{EH-action}). Since we have $\hat{W}=0$ in~(\ref{LambdaW}) and $p_{\text{max}}=0$ in~(\ref{rho}) for the gauge reducer~(\ref{rho-LG}), in the decomposition~(\ref{Jmu-3}) both the on-shell vanishing part $\Psi^\mu$~(\ref{Psimu}) and the conditionally trivial part $\Theta^\mu$~(\ref{Thetamu}) vanish identically, $\Psi^\mu=0$, $\Theta^\mu=0$. Hence, from~(\ref{Jmu}),~(\ref{J-2}) we obtain
\begin{equation}\label{Jmu-LG}
\displaystyle J^\mu=\Omega^\mu=\stackrel{(1)}J\!^\mu_\lambda\omega^\lambda+\stackrel{(2)}J\!^{\mu\nu}_\lambda\partial_\nu\omega^\lambda=\partial_\nu\Big(\!\stackrel{(2)}J\!^{\mu\nu}_\lambda\omega^\lambda\Big)\,, \qquad \partial_\mu J^\mu\equiv 0\,,
\end{equation}
where $\displaystyle \stackrel{(1)}J\!^\mu_\lambda=-\,2\eta_{\lambda\nu}{\delta L}/{\delta h^{\mu\nu}}$, and $\displaystyle \stackrel{(2)}J\!^{\mu\nu}_\lambda=\,\stackrel{(2)}\Omega\!^{\mu\nu}_\lambda=-\!\stackrel{(2)}\Omega\!^{\nu\mu}_\lambda$ (see~(\ref{J-2}),~(\ref{Omegamu})) takes the form
\begin{equation}
\displaystyle \stackrel{(2)}J\!^{\mu\nu}_\lambda=\partial^\nu h^\mu{}_\lambda-\partial^\mu h^\nu{}_\lambda+\delta^\mu_\lambda(\partial_\rho h^{\nu\rho}-\partial^\nu h)-\delta^\nu_\lambda(\partial_\rho h^{\mu\rho}-\partial^\mu h)\,.
\end{equation}
Thus, the current~(\ref{Jmu-LG}) is trivial (improper) because it reduces entirely to the divergence of a skew-symmetric tensor. Its divergence vanishes identically off-shell, regardless of whether the fields $h^{\mu\nu}$ obey the equations of conditional extremum~(\ref{Field-Eq-LG}) or the constraining equations~(\ref{Cotton}).

\section{Concluding remarks}
Let us briefly summarize the main results of the article and discuss the open issues.

We consider field theories in which the admissible field configurations are constrained by differential equations from the outset. 
This setup implies that the action must be minimized within the class of admissible fields. 
In this conditional extremum problem, we do not introduce Lagrange multipliers or any other auxiliary fields to avoid the inclusion of additional degrees of freedom.
Instead, we use the gauge symmetry of the equations imposed to constrain the admissible fields.  We exploit the fact that such a gauge variation of the action must vanish at its conditional extremals~\cite{Lyakhovich:2024yai}. This leads to a self-contained system of equations for the original fields (\ref{Field-Eq}), which any conditional extremal of the action must satisfy.

The main issue addressed in this article is the structure of the gauge symmetry of the conditional extremum problem, and the connection between gauge symmetry and gauge identities in this class of field theories. The gauge symmetry of a complete conditionally Lagrangian system is a reduction of the gauge symmetry of the system of equations imposed to constrain the admissible fields. We introduce a generating set of operators that explicitly implement the reduction of a broader symmetry of the constraining equations to a narrower symmetry of the equations for conditional extremals of the action. We call these operators gauge symmetry reducers.  We demonstrate that the gauge symmetry of the complete conditionally Lagrangian system does not necessarily respect the Lagrangian even in the class of admissible fields. There is an ambiguity in the Lagrangian which does not have an immediate analogue for the unconditional extremum problem. The proper gauge symmetry is defined to respect both the constraints and the Lagrangian. Moreover,  the proper gauge symmetry of any conditionally Lagrangian system is in one-to-one correspondence with the Noether identities between the field equations. This can be considered an extension of the second Noether theorem and its converse to the conditional extremum problem. The Noether identities are generated by the formal Hermitian conjugates of the gauge symmetry reducers. Trivial gauge symmetry reducers can lead to trivial identities that involve only the constraining equations and hold regardless of the action. We also consider the issue of conserved currents corresponding to the Noether identities and demonstrate that they are improper in the sense that they can always be decomposed into three groups of terms, namely, vanishing on-shell, conditionally trivial and having an off-shell vanishing divergence.

Now, we would like to briefly comment on the issue of BRST embedding and quantization of the class of field theories whose dynamics is based on the conditional least action principle.
This class is intermediate, in a sense, between Lagrangian field theories and general non-Lagrangian dynamics.
For any system of general field equations, not necessarily Lagrangian, the classical BRST complex can be systematically constructed~\cite{Kazinski:2005eb,Lyakhovich:2004xd} at the level of the BRST differential, which is unrelated to any Poisson bracket or antibracket. The BRST differential is constructed as a nilpotent odd vector field of ghost number~1 proceeding directly from the classical field equations. This BRST operator is not a Poisson bracket with any BRST charge, nor is it an antibracket with a BV master action.  Once the BRST differential is constructed, it is sufficient for many applications of BRST theory. For example, the full-scale theory of local BRST cohomology is known~\cite{Lyakhovich:2011xy} providing a physical interpretation for the most important BRST cohomology groups, though the connections are typically less tight between these groups compared to the better-known Lagrangian case~\cite{Barnich:2000zw}.  If the local BRST cohomology of a non-Lagrangian theory has a non-trivial cohomology group associated with the Lagrange anchor~\cite{Kazinski:2005eb}, then the theory can be covariantly quantized in various ways~\cite{Kazinski:2005eb,Lyakhovich:2005mk,Lyakhovich:2006sc} that take the BRST cohomology to the quantum level. From the perspective of deformation quantization, gauge-invariant dynamics does not need to be brought into the realm of the constrained Hamiltonian formalism; it can be quantized using a more relaxed setup of weak Poisson brackets~\cite{Lyakhovich:2004xd}. The existence of a Lagrange anchor is less restrictive for the field equations than the existence of a master action that generates the BRST differential via the antibracket. Similarly, the existence of the weak Poisson bracket~\cite{Lyakhovich:2004xd} is less restrictive than the existence of a BRST charge that generates the BRST transformation as the Hamiltonian vector field. Since the conditional extremum problem involves the action from the outset, and admits a Hamiltonian formalism at least when there is no gauge symmetry~\cite{Lyakhovich:2024yai}, one can expect that the BRST complex can be constructed in a way that involves the BV master action and the BFV-BRST charge, though these structures should have essential distinctions from the well-known patterns of Lagrangian systems.  In particular, the antibracket cannot be canonical, as there is no pairing between the fields and equations, and hence fields and antifields cannot be conjugate variables in this case. The structure of the antibracket should take into account the data of the conditional extremum problem, including the constraint equations and their gauge symmetries. These issues will be addressed elsewhere.

Let us make a final remark that the system of conditionally Lagrangian field equations~(\ref{Field-Eq}), by construction, admits a Lagrange anchor~\cite{Kazinski:2005eb} defined by generators $\hat{R}$.
We do not elaborate on this observation here, leaving it for future work. The structure of the anchor  implies that the degrees of freedom governed by the constraining equations will not fluctuate upon quantization based on the corresponding methods~\cite{Kazinski:2005eb,Lyakhovich:2004xd,Lyakhovich:2005mk,Lyakhovich:2006sc}.  
In particular, the path integral for the transition amplitude, being a solution of the Schwinger-Dyson equation for non-Lagrangian systems with a Lagrange anchor~\cite{Lyakhovich:2005mk}, will be localized on field configurations for which $T_a=0$.  This opens up a somewhat speculative possibility of freezing some of the degrees of freedom at the quantum level, while allowing them to continue their classical evolution. 
  
\vspace{0.2 cm}
 
\noindent
\textbf{Acknowledgements.} We thank M.~A.~Grigoriev and A.~A.~Sharapov for discussions. 
The work is supported by the research project FSWM-2025-0007 of the Ministry of Science and Higher Education of the Russian Federation.

\end{document}